\documentclass{PoS}

\title{Current status of LEGEND: Searching for Neutrinoless Double-Beta Decay in $^{76}$Ge: Part I}

\ShortTitle{LEGEND Status}

\author{I.Guinn \thanks{Speaker of ``LEGEND: Searching for Neutrinoless Double-Beta Decay in $^{76}$Ge''.}\\
        University of North Carolina\\
        E-mail: \email{iguinn@uw.edu}}
\author{J.M.L\'opez-Casta\~no \thanks{Speaker of ``Current Status of LEGEND''.}\\
        University of South Dakota\\
        E-mail: \email{Mariano.Lopez@usd.edu}}

      \abstract{Neutrinoless double-beta decay($0\nu\beta\beta$) decay is a hypothetical process that violates lepton number, and whose observation would unambiguously indicate that neutrinos are Majorana fermions. In the standard inverted-ordering neutrino mass scenario, the minimum possible value of m$_{\beta\beta}$ corresponds to a half-life around 10$^{28}$ yr for $0\nu\beta\beta$ decay in $^{76}$Ge, which is the target of the next generation of experiments. The current limits of GERDA and \textsc{Majorana Demonstrator} indicate a half-life higher than 10$^{26}$ yr. These experiments use high-purity germanium (HPGe) detectors that are highly-enriched in $^{76}$Ge. They have achieved the best intrinsic energy resolution and the lowest background rate in the signal search region among all $0\nu\beta\beta$ experiments.
        
Taking advantage of these successes, a new international collaboration - the Large Enriched Germanium Experiment for Neutrinoless $\beta\beta$ Decay (LEGEND) - has been formed to build a ton-scale experiment with discovery potential covering the inverse-ordering neutrino mass range in a decade, following a phased approach. This first part of LEGEND proceedings describes GERDA and \textsc{Majorana Demonstrator} capabilities and the general plan of LEGEND to reach the goal, while the second part is focused in the status of the first stage of LEGEND, LEGEND-200.}

\FullConference{XXIX International Symposium on Lepton Photon Interactions at High Energies - LeptonPhoton2019\\
		August 5-10, 2019\\
		Toronto, Canada}

\begin{document}

\begin{figure}[t]
  \begin{center}
    \includegraphics[height=4.5cm]{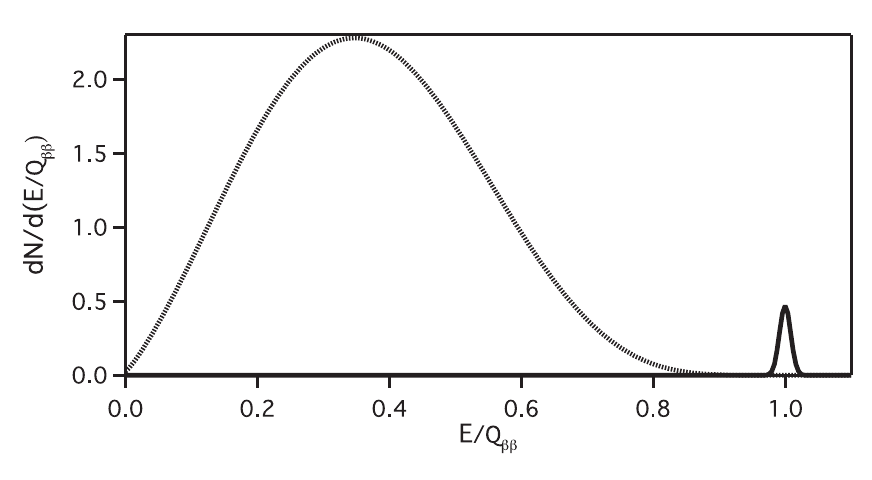}
    \includegraphics[height=4.8cm]{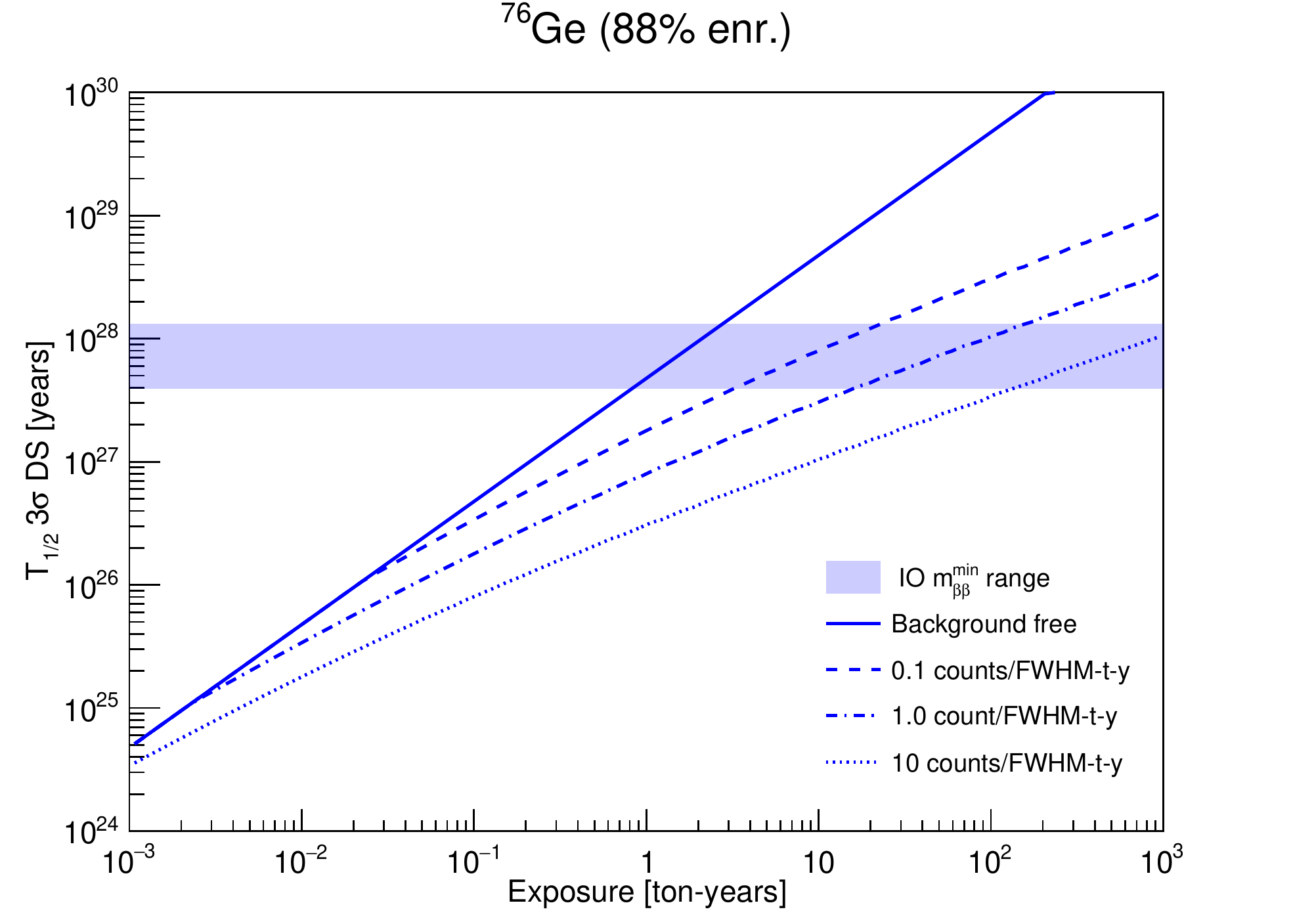}
    \caption{ 
      Left: a simulated $0\nu\beta\beta$-decay electron energy spectrum. The broad spectrum comes from $2\nu\beta\beta$ while the peak at the endpoint comes from $0\nu\beta\beta$, with a branching ratio of 1\%. 
      Right: Discovery potential at 3$\sigma$ confidence for $0\nu\beta\beta$ as a function of isotopic exposure and background index. The blue band represents the range of half-lives corresponding to the 17~meV.\cite{Steve}
      \label{Sensitivity}
    }
  \end{center}
\end{figure}

\section{Introduction}

Double beta decay (2$\nu\beta\beta$) is a process in which two neutrons change into two protons emitting two electrons and two anti-neutrinos, and has been observed in several isotopes.
Neutrinoless double beta decay ($0\nu\beta\beta$) decay is a similar process in which the anti-neutrinos are not emmited: $X^{N}_{Z} \rightarrow Y^{N}_{Z+2}+2e^{-}$.
The signature of $0\nu\beta\beta$ decay is the emission of two electrons with total energy corresponding to the decay Q-value, and its half-life ($T^{0\nu}_{1/2}$) is proportional to the effective neutrino mass ($m_{\beta\beta}$).
The goal for the next generation of $0\nu\beta\beta$ experiments is a sensitivity to $m_{\beta\beta}$ of $\sim$17~meV, which  cover all possible values in the inverted-ordering neutrino mass scenario \cite{IO}.

$^{76}$Ge undergoes $2\nu\beta\beta$, and several experiments are actively searching for $0\nu\beta\beta$ in $^{76}$Ge.
These experiments utilize High-Purity Germanium (HPGe) detectors which are enriched to at least 85\% in $^{76}$Ge, meaning the $^{76}$Ge acts as both source and detector of $0\nu\beta\beta$.
Different variations of semi-coaxial or P-type Point Contact (PPC) detector geometries are used, because they provide advantages in energy resolution and background discrimination over planar and coaxial ones.
Achieving a sensitivity to 17~meV of $m_{\beta\beta}$ requires a half-life sensitivity of $10^{28}$~y, which can be achieved with $\sim10$~tonne-y of exposure with $<0.1$~cts/FWHM-t-y of backgrounds in the region of interest for $0\nu\beta\beta$, as shown in Figure~\ref{Sensitivity}.

\subsection{GERDA}

\begin{figure}
  \begin{center}
    \includegraphics[height=3.7cm]{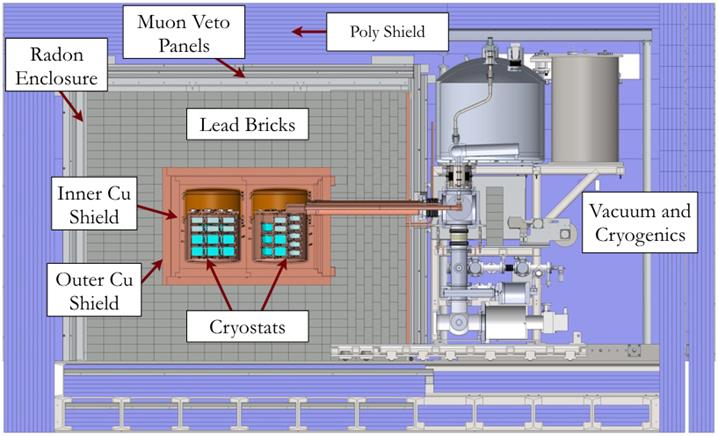}
    \includegraphics[height=3.7cm]{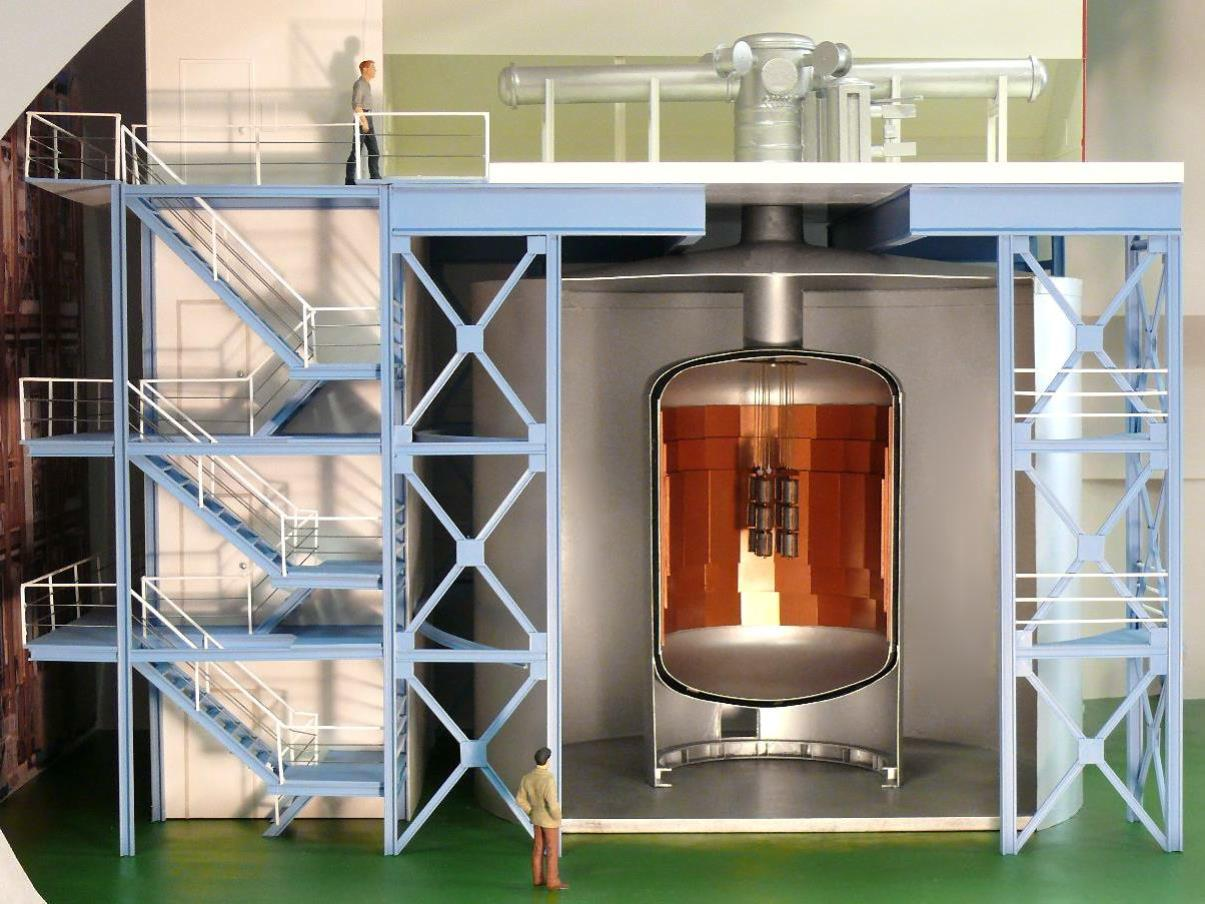}
    \includegraphics[height=3.7cm]{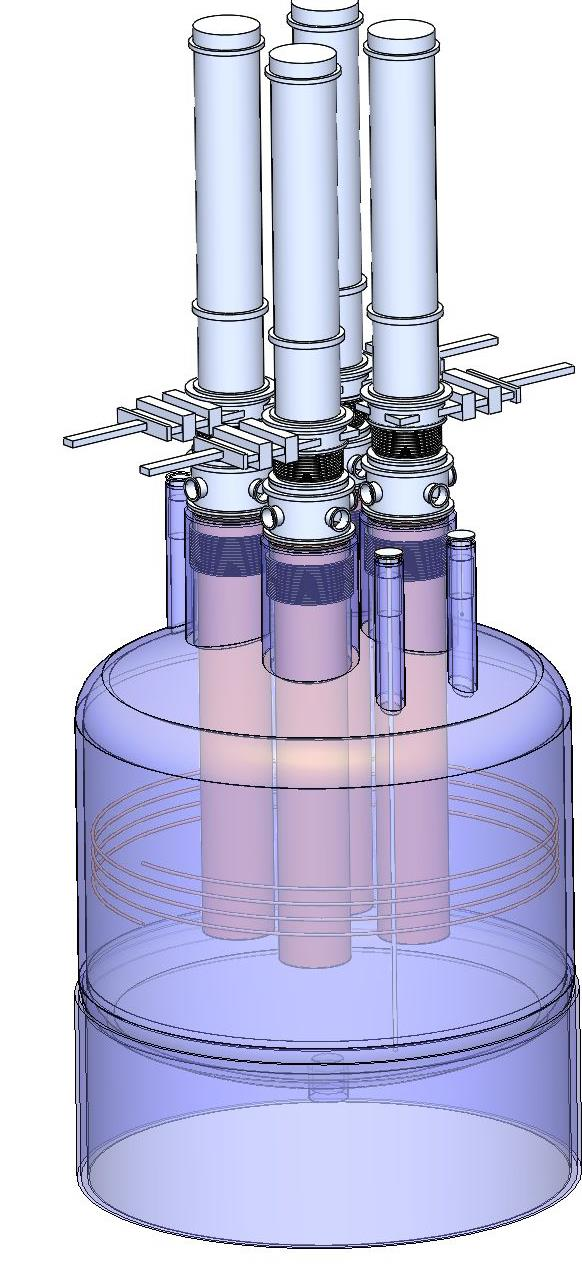}
    \caption{ 
      Left: A drawing of the \textsc{Majorana Demonstrator} experiment.
      Middle: A drawing of the GERDA experiment. The LEGEND-200 experiment is repurposing this infrastructure for a larger detector array.
      Right: A baseline design for LEGEND-1000.
      \label{Experiments}
    }
  \end{center}
\end{figure}

GERDA \cite{GERDA}, shown in Figure~\ref{Experiments}, is located in the Laboratori Nazionali del Gran Sasso (LNGS).
It utilizes 15.6~kg of semi-coaxial detectors and 20~kg of Broad Energy Germanium (BEGe) detectors.
To achieve low background, the detectors are immersed in liquid argon (LAr), which acts as a scintillating active veto.
GERDA has collected 46.7~kg-y of isotopic exposure, with a background index of 2.6~cts/FWHM-t-y for the BEGe detectors.
GERDA has achieved a median $T^{0\nu}_{1/2}$ sensitivity of $5.8\times10^{25}$~y and 90\% CL limit of $8\times10^{25}$~yr\cite{GERDA}.
These results represent the lowest background index (BI) and highest half-life sensitivity of any currently operating $0\nu\beta\beta$ experiment.

\subsection{\textsc{Majorana Demonstrator}}

The \textsc{Majorana Demonstrator} (MJD) \cite{MJD}, shown in figure~\ref{Experiments}, is operating 44.8~kg of PPC HPGe detectors in vacuum at the Sanford Underground Research Facility (SURF).
29.7~kg of the detectors are enriched to 88\% in $^{76}$Ge, while remainder are natural isotopic abundance BEGe detectors.
MJD uses ultra-clean materials in its construction, including copper that was electroformed underground to minimize cosmogenic activation and specially designed cables, connectors and low-noise electronics.
The enriched detectors have achieved a world-leading energy resolution of 2.5~keV at the 2039~keV Q-value (0.12\%).
MJD has collected 26~kg-y of isotopic exposure with a BI of 4.0~cts/FWHM-t-y, and has established a $T^{0\nu}_{1/2}$ limit of $2.6\times10^{25}$~y at 90\% CL\cite{MJD}.

\section{LEGEND}

LEGEND aims to explore a region of $0\nu\beta\beta$ half-life in $^{76}$Ge improved by 2 orders of magnitude, from 10$^{26}$~yr to 10$^{28}$~yr\cite{LEGEND}.
To achieve this, LEGEND will combine the best technologies of GERDA and the \textsc{Majorana Demonstrator}.
LEGEND will be operated in two phases.
The first phase, LEGEND-200, is currently being constructed as a 200~kg detector array with a background goal of $<0.6$~cts/FWHM-t-y and is expected to begin operation at LNGS in 2021.

LEGEND-1000 is the next phase of LEGEND, currently in the design stage, that will consist of a $\sim$1~tonne array of HPGe detectors immersed in LAr.
Active R\&D efforts are underway to achieve a further reduction in backgrounds to $<0.1$~cts/FWHM-t-y.
These include the development of larger HPGe detectors, optimized LAr light collection, improvements in clean materials and handling, and the use of depleted argon from underground sources.

\end{document}